\documentclass[usenatbib,twocolumn]{article}
\usepackage{graphicx}
\usepackage{amsmath,amssymb}
\newcommand{\lo}{{\cal L}_0}
\newcommand{\be}{\begin{equation}}
\newcommand{\ee}{\end{equation}}

\title{Dome C coherence time statistics from DIMM data}

\author{E. Aristidi,
 A. Agabi, 
L. Abe,
 E. Fossat,
 A. Ziad,
D. M\'ekarnia\\
{\small Laboratoire Lagrange, Universit\'e C\^ote d'Azur, Observatoire de la C\^ote d'Azur, CNRS, Parc Valrose, 06108 Nice Cedex 2, France}
}
\date{\scriptsize \sl MNRAS, 2020}

\begin{document}

\maketitle

\section*{abstract}
We present a reanalysis of several years of DIMM data at the site of Dome C, Antarctica, to provide measurements of the coherence time $\tau_0$. Statistics and seasonal behaviour of $\tau_0$ are given at two heights above the ground, 3m and 8m, for the wavelength $\lambda=500$nm. We found an annual median value of $2.9$ms at the height of 8m.  A few measurements could also be obtained at the height of 20m and give a median value of 6ms during the period June--September. For the first time, we provide measurements of $\tau_0$ in daytime during the summer, which appears to show the same time dependence as the seeing with a sharp maximum at 5pm local time. Exceptional values of $\tau_0$ above 10ms are met at this particular moment. The continuous slow variations of turbulence conditions during the day offers a natural test bed for a solar adaptive optics system.\\

keywords: Instrumentation: atmospheric effects, site testing -- Physical processes: turbulence -- Methods: data analysis

\section{Introduction}
\label{par:intro}
In the years 2000-2012, a long term program of site testing was conducted at the Concordia station on the site of Dome C (Antarctica) by the University C\^ote d'Azur (France) under the names of Concordiastro and Astroconcordia. A large variety of instruments were deployed, mainly to measure properties related to optical turbulence. Long-term statistics of turbulence parameters as well as vertical profiles of turbulence were obtained.  In particular, we pointed out the bimodal distribution of the seeing caused by a strong turbulent surface layer whose thickness was estimated to about 30m \cite{Aristidi09, Aristidi12, Trinquet08, Aristidi15}. 

The first instrument that we exploited at Concordia from 2002 was the Differential Image Motion Monitor (DIMM). This monitor, introduced in the early 90s \cite{SarazinRoddier90} has become a standard in seeing measurements. Using several DIMMs during 10 years, we obtained seeing statistics at 3 different elevations (3m, 8m and 20m above the ground level (AGL)). Isoplanatic angle measurements could also be done by modifying the pupil mask of the DIMM \cite{Aristidi05}, and the outer scale was measured by combining two DIMMs into a Generalized Seeing Monitor (GSM) configuration \cite{Ziad08}. However, no measurements of the coherence time $\tau_0$ were made with the DIMMs, the only published values so far come from the Australian automated observatory AASTINO \cite{Lawrence04}, balloon-borne radiosoundings \cite{Trinquet08} and the Single Star Scidar \cite{Giordano12}. All these measurements were made at night time during the period 2004--2006. The coherence time in the surface layer (up to an altitude of 45m) was also measured during several years by a set of sonic anemometers placed on the 45m high US tower \cite{Aristidi15}.
 
Recent theoretical advances \cite{Ziad12} allowed us to derive the coherence time from DIMM data. The method gives satisfactory results and is used routinely to process measurements from our Generalized DIMM (GDIMM, \cite{Aristidi19}) operated at the Calern observatory (south of France) as a part of the Calern Atmospheric Turbulence Station \cite{Ziad18}. We then decided to re-process Dome C DIMM data using this method, to extract the coherence time $\tau_0$. We present statistics for this parameter and comparisons with previous determinations. We also present the first daytime measurements of the coherence time during the summer.

The paper is organised as follows. Section~\ref{par:theory} recalls theoretical concepts about the coherence time and presents the method to derive it from DIMM data. The data processing and error analysis are presented in section~\ref{par:obs}. Statistics for $\tau_0$ at two elevations, and daytime values and behavior are shown in section~\ref{par:results}.

\section{Theoretical background}
\label{par:theory}
Atmospheric optical turbulence is described by a set of quantities such as the seeing $\epsilon$, the isoplanatic angle $\theta_0$, the coherence time $\tau_0$ or the spatial coherence outer scale ${\cal L}_0$. These parameters are sometimes denoted as ``integrated parameters'' as they result from an integration along the line of sight of local quantities (wind speed, refractive index structure constant\ldots). A review of optical turbulence in astronomy is found in \cite{Roddier81}, and we will focus here on the coherence time $\tau_0$.

The coherence time relevant for AO is defined by \cite{Roddier81}:
\be
\tau_0=0.31\, \frac{r_0}{v_e}
\label{eq:tau0}
\ee
with $r_0$ the Fried parameter and $v_e$ the effective wind speed defined as a weighted average of the wind speed $V(h)$ over the whole atmosphere:
\be
v_e=\left[\frac{\int_0^\infty |V(h)|^{5/3} \, C_n^2(h) \, dh}{\int_0^\infty C_n^2(h)\, dh}\right]^{3/5}
\label{eq:veffdef}
\ee
where $C_n^2(h)$ is the refractive index structure constant at an altitude $h$. It has been shown \cite{Conan00, Ziad12,Aristidi19} that $v_e$ can be estimated from DIMM data. It is deduced from the temporal structure functions $D_{x|y}(\tau)$ of the Angle of Arrivals (AA) in the $x$ and $y$ directions ($x$ being parallel to the right ascension and $y$ to the declination axis):
\be
D_{x|y}(\tau)=\langle (x|y(t)-x|y(t+\tau))^2 \rangle_t
\ee
 where $\langle \rangle_t$ stands for temporal average. This function is zero for $\tau = 0$ and saturates to a value $D_s$ for large $\tau$.  We define the AA correlation time $\tau_{A,x|y}$ as the time $\tau$ for which { $D_{x|y}(\tau)=\frac{D_s}{k'}$}, i.e. when the curve of is $D_{x|y}(\tau)/D_s$ is {$1/k'$} above zero, as shown in Fig.~\ref{fig:ex_struct_fn} .{The constant $k'$ is here taken as $k'=e$ to be consistent with the GSM \cite{Ziad12}.} Typical values for $\tau_{A}$ are 10 to 30ms. The effective wind speed $v_e$ is given by {eq.~5.16 of \cite{Conan00}}:
\be
v_e=10^3 D\, G^{-3} \left[\tau_{A,x}^{\frac 1 3} + \tau_{A,y}^{\frac 1 3}  \right]^{-3}
\label{eq:veff}
\ee
where $D$ is the sub-pupil diameter and $G$ a constant {given by eq.~13 of \cite{Ziad12}:
\be
G=7.30\, k'^{-1}\, K^{\frac 1 3}+7.01\, (1-k'^{-1})
\ee}
with $K=\frac{\pi D}{\lo}$ and $\lo$ the outer scale, taken equal to 7m, the median value at the Dome C site \cite{Ziad08}. 

\section{Observations and data processing}
\label{par:obs}
The data analyzed in this paper come from 3 DIMMs. Two of them were located at ground level (with a pupil height of about 3m AGL), as a part of the GSM experiment. When the two telescopes observed simultaneously, we took the mean value of $\tau_0$ given by the two instruments. The 3rd DIMM was on top of a platform, at an elevation of 8m. The dataset covers the period { Dec 2006 -- Oct 2011 for the ground DIMMs, and Dec 2005 -- Sep 2012 for the DIMM at 8m.

In 2005 and 2012, one of the ground DIMMs was put on the roof of the calm building of the Concordia station, at an elevation of 20m AGL. Data were collected for several months during the second half of the two winterovers, and allowed to measure the seeing at 20m. However, because of strong vibrations of the building and several technical failures, the number of data exploitable for coherence time measurement is small, and limited to the period June--September of the two years.

The instrumentation and the data processing for seeing estimation are described in details in \cite{Aristidi05}. The raw data consist of a series of photocenter positions for the twin images produced by the telescope. Coordinates lists are divided into short sequences of 2mn. They are pre-processed to remove bad points, drift and vibrations. For every sequence, the AA structure function is computed, the AA correlation time $\tau_A$ is estimated {and we derive the effective wind speed $v_e$}. The Fried parameter is also calculated by the classical DIMM method. This lead, {through Eq.~\ref{eq:tau0}}, to an estimation of the coherence time $\tau_0$ every 2mn. The high framerate of about 100 frames/second allows to properly sample AA structure functions and to reduce the statistical error, as several thousands of photocenter coordinates are measured in a 2mn time interval.

\begin{figure}
\includegraphics[width=90mm]{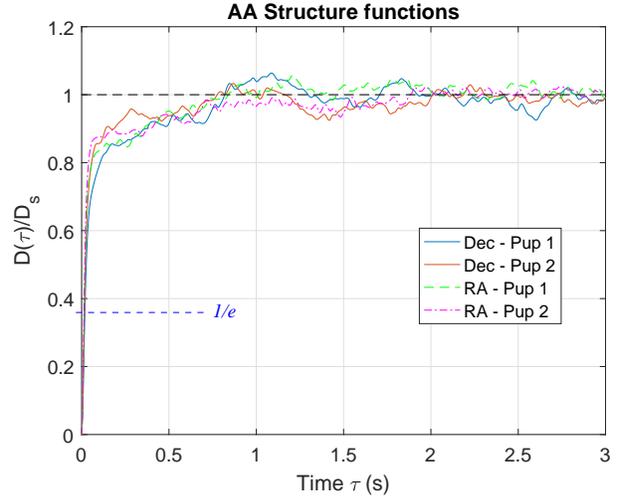}
\caption{Example of AA structure functions $D_A(\tau)$ calculated on data from the ground DIMM, on January 18th, 2011 at 0:33 local time. The 4 curves displayed correspond to $y$ (RA) and $x$ (Dec) motions for the two sub-images. One can see that all curves saturate at large temporal lag $\tau$, structure functions were divided by their saturation value $D_s$. The intersection with the line $D_A(\tau)/D_s=1/e$ (dashed blue line) gives the AA correlation time $\tau_A$. For this example, we found $\tau_{A,x1}=19$ms,  $\tau_{A,x2}=16$ms, $\tau_{A,y1}=15$ms and $\tau_{A,y2}=14$ms.}
\label{fig:ex_struct_fn}
\end{figure}

\subsection*{Error analysis}
The calculation of the error budget on the coherence time is described in section~5.3 of \cite{Aristidi19}. The uncertainty $\delta\tau$ on the AA correlation time is given by 
\be
\delta\tau=\frac{\delta D}{D'(\tau_A)}
\ee
with $\delta D$ the uncertainty on the structure function, estimated in the saturation zone as the standard deviation of values and $D'(\tau_A)$ its slope for $\tau=\tau_A$, which can be calculated as a finite difference. Errors on $\tau_A$ were computed for all 3 DIMMs for subsets of 3 months of data. We found a relative uncertainty $\frac{\delta \tau_A}{\tau_A}\simeq 7\%$ for the Dec axis and 10\% for the RA axis. These values propagate into the effective wind speed, whose uncertainty, obtained by differentiation of Eq.~\ref{eq:veff}, is $\frac{\delta v_e}{v_e}\simeq 8\%$.

Finally, the relative error on the coherence time is obtained from Eq.~\ref{eq:tau0}, assuming that errors on $r_0$ and $v_e$ are uncorrelated: 
\be
\frac{\delta \tau_0}{\tau_0}=\left[\left(\frac{\delta v_e}{v_e}\right)^2 \: + \; \left(\frac{\delta r_0}{r_0}\right)^2\right]^{\frac 1 2} \; \simeq \; 8\%
\ee

where we used a relative error on the Fried parameter $r_0$ of $\simeq 1\%$ \cite{Aristidi05}.


\begin{table*}
\begin{tabular}{lrrrrrrc}\hline
Elevation	   &   & { 3 m}   &   &   &   { 8 m}   &   & { 20m}\\ 
    &   summer    &    winter   &   total   &   summer   &   winter   &   total   &   { June--Sep}\\ \hline
Nb of points      &   73042   &   148570   &   437986   &   62126   &   65378   &   262951 &   15213\\
Median      &   5.1   &   1.3   &   1.9   &   8.2   &   1.7   &   2.9 &   6\\ 
Mean     &   6.3   &   1.8   &   3.1   &   9.9   &   2.4   &   5.2   &   7\\
1st quartile      &   3.2   &   1.0   &   1.3   &   5.0   &   1.2   &   1.6 &   3 \\
3rd quartile     &   7.8   &   1.8   &   3.2  &   13   &   2.5   &   6.0   &   8\\
1st centile     &   1.4   &   0.50   &   0.55   &   1.5   &   0.57   &   0.63  &   1 \\
Last centile     &   25   &   6.8   &   15  &   32   &   11   &   24   &   35\\ \hline
\end{tabular}
\caption{Statistics of coherence time $\tau_0$ calculated at the wavelength $\lambda=500$nm. Values are in milli-seconds}
\label{table:stat}
\end{table*}
\section{Results}
\label{par:results}
\subsection{Statistical analysis}
A total amount of about $700\, 000$ values of the coherence time were obtained { from the DIMMs at heights 3m and 8m} during the considered period. This large number of measurements allows to undertake significant statistical studies. {For the 20m DIMM, the number of values is much smaller, about 15\,000, and limited to the period June--September.}

Table~\ref{table:stat} presents statistics of $\tau_0$ obtained at { the 3 heights 3m, 8m and 20m}. All values are calculated for the wavelength $\lambda=500$nm. { For the 3m and 8m DIMMs,} global and seasonal values are given for the summer (December -- January) and the winter (June -- August). Very different conditions are observed between these two seasons: best conditions are met during the summer with a median value of 8.2ms measured by the 8m DIMM, while it is 1.7ms in the middle of the winter. The coherence time follows the same seasonal dependence as the seeing \cite{Aristidi09} and this was expected since it is proportional to the Fried parameter (Eq.~\ref{eq:tau0}).

Fig.~\ref{fig:tau0_vs_month} shows the monthly median values of $\tau_0$ at the two heights (3m and 8m). The best period is December--January, the closest to the summer solstice: this is why we chose to denote this period as ``summer'' in this paper. In winter, bad conditions are explained by the presence of the turbulent surface layer. This surface layer was observed at many locations of the Antarctic plateau \cite{Marks99, Bonner10, Okita13} and is predicted by climatic models \cite{SwainGallee06}. Its consequence is a poor winter seeing around 1.7$''$ at the height of 8m \cite{Aristidi12} and a poor coherence time (1.7ms at 8m).

Cumulative histograms of $\tau_0$ in winter and summer are presented in Fig.~\ref{fig:cdf_2alt_hiv_ete} for both altitudes 3m and 8m. In winter, the two distributions are close to each other with small difference between the two altitudes (the median values differ by 0.3ms). This difference becomes larger for large values of $\tau_0$. Hence, the 9$^{\rm th}$ decile, in winter, is 5ms at 8m and 2.5ms at 3m. This may be due to situations where the thickness of the surface layer is lower than the altitude of the 8m DIMM, which is then in the free atmosphere (FA). Such situations, mentionned in \cite{Aristidi09} and~\cite{Fossat10} are likely to occur 12\% of the time in winter \cite{Aristidi12}. { The situation is more favorable at 20m height, the DIMM being above the surface layer 45\% of the time. We found a median value of $\tau_0$ of 6ms at this altitude.} Summer curves show larger values of $\tau_0$ and will be discussed in Sect.~\ref{par:summer}.

Table \ref{table:compar} compares median annual values of $\tau_0$ obtained by the DIMMs with other available Dome C measurements. The balloon value is based on 34 radiosoundings in 2005. But part of these flights suffers from a bias due to a lack of measurement of the wind speed near the ground and/or a poor vertical sampling of the surface layer. The AASTINO value was measured by a MASS and gives the coherence time in the FA (indeed, it compares well with the FA value of 6.8ms measured by the ballons above 33m \cite{Trinquet08} {and it is consistent with our value of 6ms at the height of 20m}). The SSS value is based on a large set of vertical profiles obtained in 2006 and the instrument was at the same height as the 8m DIMM. The SSS estimation of $\tau_0$ is a little more optimistic than the DIMM one, but the small diameter of the telescope (40cm) prevents the measure of the turbulence in layers where the wind speed is higher than $\sim 40$m/s. This happens in particular at high altitudes during the winter where strong stratospheric winds are triggered by the polar vortex \cite{Giordano12, Aristidi_wind05}. The coherence time is dependent on both $C_n^2(h)$ and $V(h)$ profiles (Eqs.~\ref{eq:tau0}, \ref{eq:veffdef}) and some values of $\tau_0$ may be overestimated by the SSS.

The table \ref{table:compar} compares also the Dome C coherence time with other sites in the world. { The effective height of these measurements is about 2m AGL.} The South Pole value is derived from 15 radiosoundings in the winter 1995. {
It appears to be similar to the winter $\tau_0$ measured by our DIMMs, but the statistics is too weak to derive reliable conclusions}. The two other values at Paranal and Mauna Kea were obtained by the Generalized Seeing Monitor (GSM) \cite{Ziad00} using the same method as the present work. A more complete list of locations measured by the GSM (including La Silla, Palomar, La Palma\ldots) can be found in \cite{Ziad12}, they show values of $\tau_0$ in the range 1.2~--~6.6~ms. They suggest that the Dome C coherence time at the height of 8m AGL is comparable to the one observed above temperate astronomical sites. { At the height 20m however, Dome~C becomes better that the majority of sites.}

\begin{figure}
\includegraphics[width=90mm]{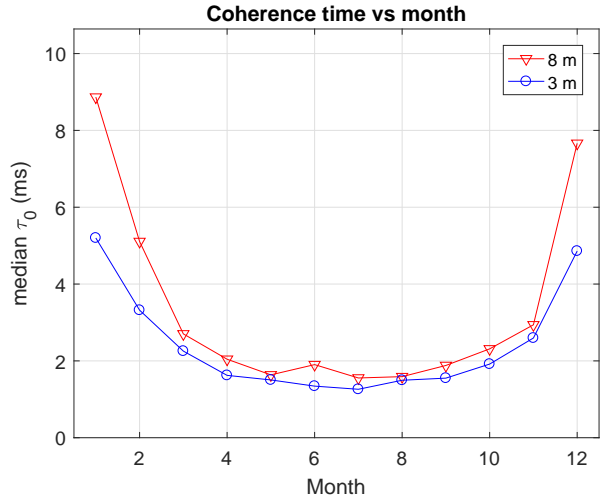}
\caption{Monthly median values of $\tau_0$ at the two altitudes 3m and 8m.}
\label{fig:tau0_vs_month}
\end{figure}

\begin{figure}
\includegraphics[width=90mm]{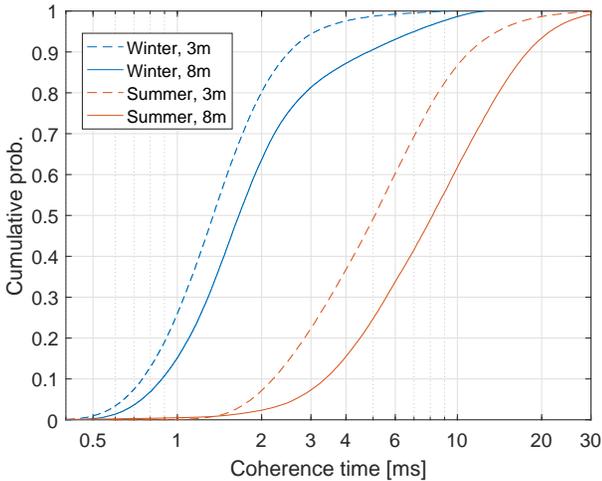}
\caption{Cumulative distributions of the coherence time at altitudes 3m and 8m, during the summer (Dec--Jan) and the winter (June--Aug)
}
   \label{fig:cdf_2alt_hiv_ete}
\end{figure}

\begin{table}
\begin{tabular}{lccl}\hline
        & $\tau_0$ (ms) & Height & Reference\\ \hline
DIMM 3m & 1.9 & 3m& \\
DIMM 8m & 2.9 & 8m & \\
{DIMM 20m} & 6 & 20m & \\
SSS & 3.4 & 8m &\cite{Giordano12}  \\
Balloons & 5.7 & 8m& \cite{Trinquet08}  \\
AASTINO & 7.9 & 30m& \cite{Lawrence04}   \\ \hline
South Pole & 1.6 & 2m & \cite{Marks99}  \\
Paranal & 3.2 & 2m & \cite{Ziad12}\\
Mauna Kea & 2.4 & 2m & \cite{Ziad12} \\ \hline
\end{tabular}\caption{Coherence time median values from DIMMs at altitudes 3m, 8m { and 20m}, compared to other Dome C mesurements and to other sites. The column 2 reports the effective height above which measurements were done. { For the 20m DIMM, the statistics is poor and measurements are available only during the period June--September.}}
\label{table:compar}
\end{table}

\subsection{The summer situation}
\label{par:summer}
The site of Dome C offers very particular turbulence conditions during the summer. We have discovered \cite{Aristidi05, Aristidi09} that every day, the planetary boundary layer (responsible for more than 90\% of the optical turbulence) almost completely vanishes around 5pm local time, due to the cancellation of the temperature gradient between the ground and the upper atmosphere. Excellent seeing values around 0.4$''$ are observed during this transition from the day to the night (though there is no night is summer, but a period where the Sun in low above the horizon). Similar conditions were reported at other locations of the Antarctic plateau, such as Dome~F \cite{Okita13} and Taishan station \cite{Tian20}. The coherence time being proportional to the Fried parameter (Eq~\ref{eq:tau0}), it was natural to suspect that $\tau_0$ might be very high at this particular moment, and this was comforted by visual observations at the telescope showing very slow image motions. However we did not have quantitative measurements at the time of the observations. 

The present dataset contains 6 summer campaigns, at heights 3m and 8m. Statistics for the summer are presented in Table~\ref{table:stat}. From the cumulative distributions displayed in Fig.~\ref{fig:cdf_2alt_hiv_ete}, it can be seen that the summer coherence time at 8m is very high, with values greater than 10ms 38\% of the time (13\% at 3m).

Fig.~\ref{fig:tau0_vs_time_ete} shows the hourly median value of $\tau_0$ as a function of the local time. For the 8m DIMM, values around 5ms are observed at night hours, and the coherence time progressively increases at the end of the morning, to attain a maximum near 12ms at 17h local time. This peak is coincident with the seeing minimum observed every day in mid-afternoon in summer (fig.~6 of \cite{Aristidi05}). 

This particular period around 5pm gives an opportunity to infer statistical properties of the coherence time in the FA. This statistics, made possible by the large amount of data, is presented in Fig.~\ref{fig:histo_tau0}. More than 7000 values of $\tau_0$ were gathered by the 8m DIMM between 4pm and 6pm during the months of December and January. Their histograms shows a log-normal distribution of median 12ms. Note that this summer FA coherence time is greater than FA measurements given by the balloons (6.8ms, \cite{Trinquet08}) and the MASS (7.9ms, \cite{Lawrence04}). The difference is explained by stronger high-altitude wind speed in the winter.

\begin{figure}
\includegraphics[width=90mm]{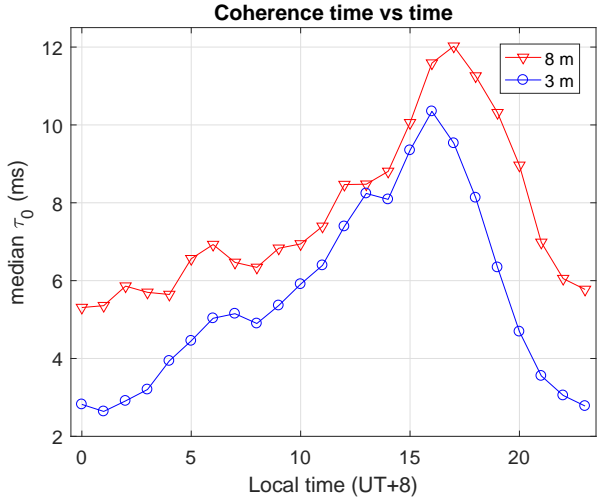}
\caption{Hourly median values of $\tau_0$ in the Summer (Dec--Jan) the two altitudes 3m and 8m.}
\label{fig:tau0_vs_time_ete}
\end{figure}

Table \ref{table:stat} shows that the coherence time, for the 8m DIMM, can be greater than 32ms in 1\% of cases. Indeed, superb conditions were sometimes met as in the case of Fig.~\ref{fig:seeingtsgood}, obtained in the afternoon of Dec, 25th, 2009. The coherence time remained continuously above 20ms between 4pm and 7pm, and attained values close to 50ms at 5pm.

\begin{figure}
\parbox[c]{55mm}{
\includegraphics[width=55mm]{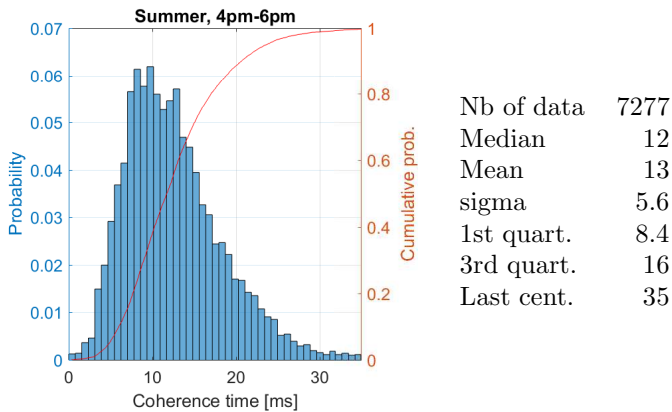}}\ \ \ 
\parbox[c]{25mm}
{\begin{tabular}{lr}
Nb of data & 7277\\
Median & 12\\
Mean    &   13  \\ 
sigma   &   5.6   \\ 
1st quart.  &   8.4   \\ 
3rd quart.  &   16 \\   
Last cent. &  35    
\end{tabular}
}
\caption{Histogram of $\tau_0$ between 4pm and 6pm during the Summer (Dec--Jan), for the DIMM at altitude 8m.}
\label{fig:histo_tau0}
\end{figure}

\begin{figure}
\includegraphics[width=80mm]{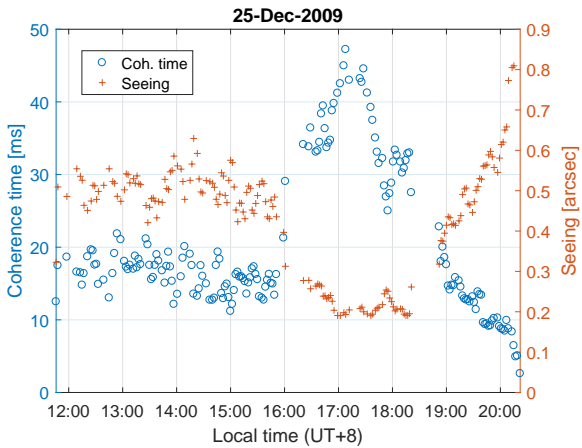}
\caption{Exceptional conditions observed on Dec. 25th, 2009 during the afternoon: the coherence time, measured by the 8m DIMM, attained values close to 50ms at 5pm local time.}
   \label{fig:seeingtsgood}
\end{figure}

\section{Conclusion}
We have presented statistics for the coherence time at Dome C, during the summer and the winter, at {three} elevations above the ground. They were obtained by reanalysis of a large dataset of DIMM data covering several years, using recent algorithms. This analysis was made possible thanks to the fast temporal sampling of DIMM images. The presence of data obtained during the daytime allowed, for the first time, to measure the coherence time at Dome~C in the summer.

As expected the coherence time { near the ground} is rather poor in winter because of the strong surface layer which contains about 90\% of the turbulence. This surface layer is much less active in summer and the coherence time is multiplied by 5 at the altitude of 8m. Indeed, in the past, we measured the coherence time $\tau_{0,s}$ inside the surface layer (between 8m and 45m AGL) with a set of sonic anemometers placed on a tower. Fig.~15 of \cite{Aristidi15} shows that values of $\tau_{0,s}$ vary from 5--6ms in winter to 30ms in summer: this is also a factor 5. Also, the surface layer is rather thin and we have reported that a telescope placed at a height of 20m would be in the free atmosphere almost 50\% of the time \cite{Aristidi12}. { Measurements made by the 20m DIMM showed very good coherence time values with a median of 6ms.}
}

The behaviour of $\tau_0$ with time during the summer is another interesting result of this work. As for the seeing, very favorable conditions are met every day around 5pm local time. Superb values greater than 12ms can be measured 50\% of the time during this period: this fraction is less than 10\% for Mauna Kea and Paranal at night \cite{Ziad12}. Combined to low seeing values around 0.4~arcsec, this offers interesting perspective for high angular resolution solar imaging and solar adaptive optics (AO). Indeed a large coherence time reduces the delay error of an AO system, while a small seeing value allows to benefit from a good correction \cite{Cheng09, Carbillet17}.

Solar adaptive optics is a well developed tool for investigating solar surface structures such as faculas, sunspots and solar granulation \cite{Rimmele11}. Solar astronomers have already expressed their interest for a solar AO instrument at Dome C \cite{Dame10}, recognizing that Dome C is a privileged site for solar observations. Indeed, Dome C summer conditions offer a unique test bench for adaptive optics. The continuous slow improvement, then degradation, of atmospheric turbulence conditions during the day offers a natural testing laboratory for an AO system. 

\section*{Acknowledgements}
The authors wish to thank the French and Italian Polar Institutes, IPEV and ENEA and the CNRS for the financial and logistical support of this programme. Thanks are also due to the Dome C local staff and people who participated to summer campaigns and winter-overs from 2005 to 2012. And we thank our referee, Pr. M. Ashley for his constructive comments which improved the paper.
\\

\bibliography{biblio1} 
\bibliographystyle{spiebib}

\end{document}